\documentclass[prb,twocolumn,showpacs]{revtex4}
\usepackage{times}

\newcommand{\be}{\begin{equation}}
\newcommand{\ee}{\end{equation}}
\newcommand{\bea}{\begin{eqnarray}}
\newcommand{\eea}{\end{eqnarray}}
\newcommand{\rd}{\mbox{d}}
\newcommand{\p}{\partial}

\newcommand{\textbinm}[2]{{\textstyle {#1\choose #2}}}
\newcommand{\abbr}[1]{{\rm #1}}
\newcommand{\eun}{{\rm e}}
\newcommand{\imu}{{\rm i}}

\usepackage{graphicx}
\usepackage{dcolumn}
\usepackage{bm}

\begin{document}

\pacs{73.43.Jn, 71.35.Lk, 71.23.An}

\title{Edge State Tunneling in a Split Hall Bar Model}

\author{Emiliano Papa and A.H. MacDonald}

\address{Department of Physics, The University of Texas, Austin, TX 78712}

\date{\today}

\begin{abstract}

In this paper we introduce and study the correlation functions of a chiral one-dimensional 
electron model intended to qualitatively represent narrow Hall bars separated into left and right sections by a 
penetrable barrier.  The model has two parameters representing respectively interactions between
top and bottom edges of the Hall bar and interactions between the edges on opposite sides of the barrier.
We show that the scaling dimensions of tunneling processes depend on the relative 
strengths of the interactions, with repulsive interactions across the Hall bar tending to 
make breaks in the barrier irrelevant.
The model can be solved analytically and is characterized by a difference   
between the dynamics of even and odd Fourier components. 
We address its experimental relevance by comparing its predictions with those of a more 
geometrically realistic model that must be solved numerically.

\end{abstract}

\maketitle

\section{Introduction}

\begin{figure}
 \unitlength=1mm
\begin{center}
\begin{picture}(70,50)
\put(-3,47){\includegraphics[width=68\unitlength,height=58\unitlength,angle=-90]{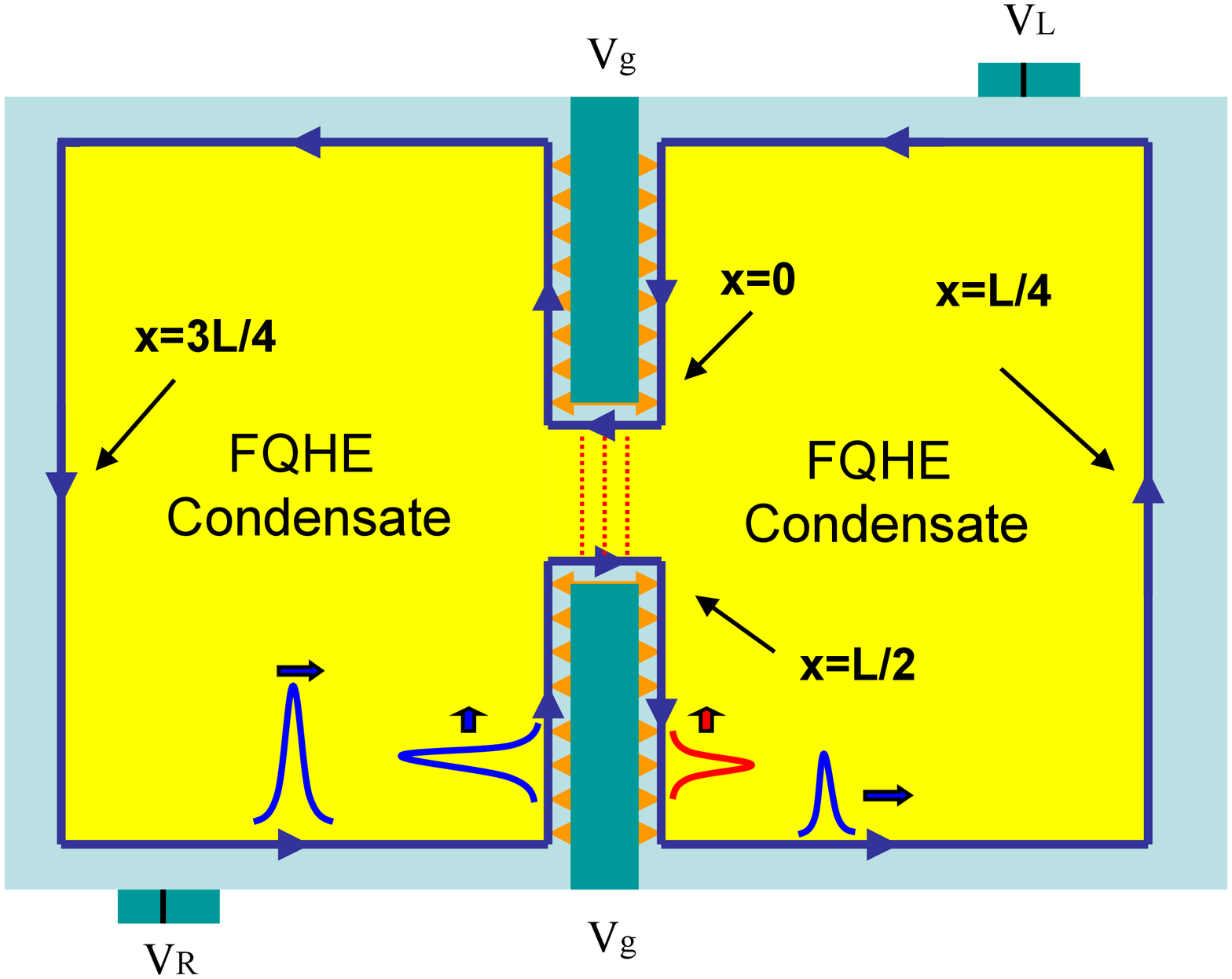}}
\end{picture}
\end{center}
\caption{Schematic illustration of a Hall bar with a split-gate constriction or a cleaved 
edge overgrowth barrier with a break.  In these geometries, a barrier separates the Hall 
bar into left and right segments.  In the split gate geometry, the constriction is created 
in order to enhance tunneling between the top and the bottom of the Hall bar.  In the cleaved
edge overgrowth case, the barrier is intended to block transport between left and right 
but inevitably has weak spots at which tunneling across has a larger bare amplitude.  In both 
cases, interactions across the line junction have an influence on transport that can compete 
with the influence of the bulk filling factor $\nu$ and the influence of interactions between
top and bottom of the Hall bar.  The split Hall bar model attempts to capture these competing 
influences qualitatively.  One consequence of interactions across the Hall bar is illustrated 
in this schematic figure; because of repulsive interactions
across the junction separating left and right hand sides of the Hall bar,
an electron approaching the constriction point will drag a fractional hole charge  
along the opposite side of the barrier.  To conserve charge on the right hand side of the Hall 
bar, a fraction of an electron charge must be simultaneously emitted along the lower edge (Ref.~\onlinecite{safi}).
}
\label{fig:one}
\end{figure}

A two-dimensional electron system on a quantum Hall (QH) plateau,
has low-energy chiral edge excitations\cite{wen} that provide a rich realization 
of one-dimensional electron physics recently reviewed by Chang\cite{changreview}.
Our interest in this paper is in quantum Hall edge transport 
experiments that can be used to study edge correlations 
by measuring low-temperature, low-bias voltage resistances
to probe the infrared scaling of weak tunneling processes between 
prescribed points on the sample edges.  Typically the tunneling amplitude 
between top and bottom edges of a Hall bar is enhanced by creating a 
constriction with gates \cite{pellegrini,heiblum,saclay,kane,fradkin,zulicke0,fendley,vignale} 
or by growing a cleaved edge overgrowth barrier\cite{kang,grayson,mitra}, 
ultimately splitting the quantum Hall liquid into two separate systems separated by a thin barrier
as illustrated schematically in Fig.~\ref{fig:one}. 
Tunneling between top and bottom of the Hall bar is enhanced by the barrier, whatever its origin.
These experiments sense the power law character of the edge Green's functions and the 
related power law suppression of the densities of states \cite{grayson2} that are 
typical of one-dimensional systems.
  
\begin{figure}      
\begin{center}
\unitlength=1mm
\begin{picture}(80,50)
\put(2,-5){\includegraphics[width=70\unitlength,height=55\unitlength]{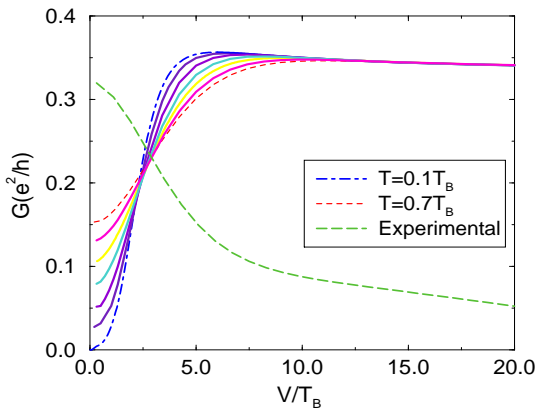}} 
\end{picture}
\end{center}
\caption{
Differential conductance across the constriction for finite temperatures, the solid lines 
ranging in incremental order with $\Delta T = 0.1T_B$ 
from $T=0.1T_B$, dotted-dashed line,  to $T=0.7T_B$, dotted line, 
obtained by numerical solution of the Bethe ansatz equations,
following for example Ref.~\onlinecite{fendley},
for bulk filling factor $\nu=1/3$ neglecting non-local interactions. 
The dashed line schematically illustrates typical experimental observations
of the differential conductance across split gate constrictions 
for $\nu=1/3$ when the gate voltage is small in Ref.~\onlinecite{pellegrini}. 
}
\label{Massless_SG_T}
\end{figure}

The present work is motivated in part by recent experiments of Roddaro
{\it et al.} 
\cite{pellegrini}
which draw attention to aspects of the transport experiments that appear to be 
inconsistent with commonly applied theoretical models\cite{kane} (see Fig.~\ref{Massless_SG_T}).
The samples that they used had length and width of comparable size $\sim 100\mu m$. 
The width of the split gate and the size of its point-contact opening are 
$\sim 100$ times smaller, suggesting an alternate description of the transport 
experiments\cite{fisher,renn} in which left and right sides of the line junction
(rather than the top and bottom of the Hall bar) are mapped to the left and right going
states of a one-dimensional electron gas. 
This geometry is rather different from what is assumed in canonical theoretical calculations  
\cite{kane,fradkin,fendley} of backscattering on Hall bars and certainly plays a role in the systematics of the
measured $I$-$V$ characteristics.  The measurements performed over a broad range of filling fractions $\nu \le 1$ 
often show a suppression of top to bottom quasiparticle tunneling at small source-drain (SD) bias voltages, 
whereas simple Hall bar models predict universal enhancement of inter-edge tunneling of fractionally charged
quasiparticles.  In general, both interactions across the line junction and interactions across the 
Hall bar can play a role in determining the $I$-$V$ characteristics of this system.  This is the feature of the 
experimental system that we attempt to capture in the split Hall bar model explained below.  

The model
we study is complementary to one analyzed by Pryadko, Shimshoni and Auerbach\cite{pryadko}
in which counter propagating edge channels approach each other at a variable angle and
interact via long-range Coulomb interactions. Our model is analytically solvable and 
captures key aspects of the geometry of cleaved-edge overgrowth line junction and
split-gate point-contact systems.  The model emphasizes the fact that both the shape 
of incompressible region edges and the locations of the points between which tunneling 
occurs can have an influence on the relevance of backscattering processes.  We refer to 
this model as the {\em split Hall bar} model.

The outline of this paper is as follows. In Sec. II we introduce the model. 
In Sec. III we discuss its bosonic and quasiparticle tunneling correlation functions.
These can then be used to evaluate the model's $I$-$V$ characteristics at finite temperature and finite size.
Finally in Sec. IV we discuss our results, comparing some predictions of this model with the 
predictions of more realistic models that must be solved numerically.
We also discuss the possible roles of edge reconstruction and 
of the incompressible strip formation \cite{chklovskii,zulicke} on the properties of 
experimental systems.

\section{The split Hall bar model}

We assume in this section that the low energy physics of a quantum Hall edge system can be expressed 
in terms of the edge charge density alone and that each chiral edge has a single channel, postponing
a discussion of the many relevant caveats 
to Sec. IV. The Hamiltonian that describes the energetics of edge fluctuations of a 
singly connected incompressible region then has the following form
\bea
H = \frac{1}{2}\int_0^L \rd x \int_0^L \rd x' \rho(x) V(x,x')\rho(x')
\label{classical_fluct}
\eea
where $x$, $x'$ are coordinates along the edges of length $L$. 
For a circular compressible region, translational invariance along the 
edge would imply that $V(x,x')$ depends only on $|x-x'|$.  Typical experimental 
geometries differ qualitatively from a circle, however, and the edge system 
is very far from translationally invariant.  The split Hall bar model attempts to
capture the most essential aspects of typical experimental geometries, without 
sacrificing the analytic solution that is enabled for this quadratic
Hamiltonian by translational invariance.   

Edge state Hamiltonians can be quantized by recognizing that charge fluctuations result from particle-hole excitations
at the edges of the Hall bar.  The following commutation relations \cite{wen} apply for particle-hole
excitations at a chiral edge: 
\bea
[\rho(x),\rho(x')]=-\frac{i\nu}{2\pi}\p_x\delta(x-x') \quad .
\label{comm_relat}
\eea
Because of the form of Eq.~(\ref{comm_relat}), it is convenient to introduce a new bosonic field\cite{Tsvelik}
related to the density by $\rho(x) = - \p_x\phi(x) / \sqrt{\pi}$,
satisfying the following commutation relations
\bea
\quad \frac{1}{\pi}[\p_x\phi(x),\p_{x'}\phi(x')] &=& -\frac{i\nu}{2\pi}
\p_x\delta(x-x') \; .
\label{charge_density}
\eea
Eq.~(\ref{charge_density}) identifies $\phi(x)$ and $\p_{x}\phi(x)$ as canonically 
conjugate variables 
$
[\phi(x),\p_{x'}\phi(x')] = - (i\nu/2)\delta(x-x')
$.
Edge fermion creation operators can be decomposed as a product
$\psi(x)=\exp\{ik_F x\} R(x)$ with 
\bea
R(x) = \frac{1}{\sqrt{2\pi}} e^{ -i\frac{\sqrt{4\pi}}{\nu}\phi(x)}
       \label{fermion}
\quad .
\eea
The analog of Eq.~(\ref{fermion}) for quasiparticles of charge $e^* = \nu e$ is
$R_{\rm QP}(x) \sim \exp\{-i\sqrt{4\pi}\phi_R(x)\}$.

The action of the split Hall bar model is therefore 
\bea
S=\frac{2i}{\nu}\int_0^L\rd x \int\rd \tau \p_x\phi(x,\tau) \p_\tau \phi(x,\tau) +\int \rd \tau H
\, ,
\label{action0}
\eea 
where $H = H_0 + H_1 + H_2$ and
\begin{equation}
H_0 = -i v_F \int_{0}^L \rd x R_{\rm QP}^\dagger(x)\p_x R_{\rm QP}(x) = \pi v_F \int_{0}^L \rd x \rho^2(x) \;.
\end{equation}
This term represents the microscopic exchange and Coulomb interactions locally at a given
point at the edge, and would be the only term if all points along the edge were equivalent.
$H_1$ and $H_2$ are intended to represent the fact that interactions between points that are 
remote, when distance is measured along the edge, can be important if the points are either 
on opposite sides of the line junction or on opposite sides of the Hall bar.  
The form we choose for these interactions is idealized in a way which yields a 
solvable model.   
To be precise we represent interactions across the constriction and interactions across the Hall bar 
by 
\bea
H_1 &=& g_1 \pi v_F \int_{0}^{L} \rd x \rho(x) \rho(L-x) \; , \;
\\[3mm]
H_2  &=& g_2 \pi v_F \int_{0}^{L/2} \rd x \rho(x) \rho(L/2-x)   \,,
\label{H_int1}
\eea
where $g_1$ and $g_2$ are dimensionless interaction parameters. The $g_1$ parameter
represents the relative strength of interactions across the constriction, 
in the horizontal direction in Fig.~\ref{fig:two}.
The $g_2$ interaction parameter accounts for interactions across the Hall bar,
the vertical direction in Fig.~\ref{fig:two}. 

To solve this model we 
adopt the following Fourier transform convention for the field $\phi$: 
\bea
\phi(x,t)=\frac{1}{L}\sum_{i}\int_{-\infty}^{+\infty} \frac{\rd \omega }{2\pi}\frac{e^{i(q_ix - \omega t)}}{\sqrt{2|q_i|}}
\phi(q_i,\omega) \quad .
\label{field1}
\eea 
where $q_i=2\pi i/L$.  
Substituting Eq.~(\ref{field1}) into Eqs.~(\ref{action0}) and (\ref{H_int1}) we obtain the following action for our
split Hall bar model:
 \begin{figure}
 \unitlength=1mm
\begin{center}
 \begin{picture}(80,50)
 \put(-4,-2){\includegraphics[height=50\unitlength,width=85\unitlength]{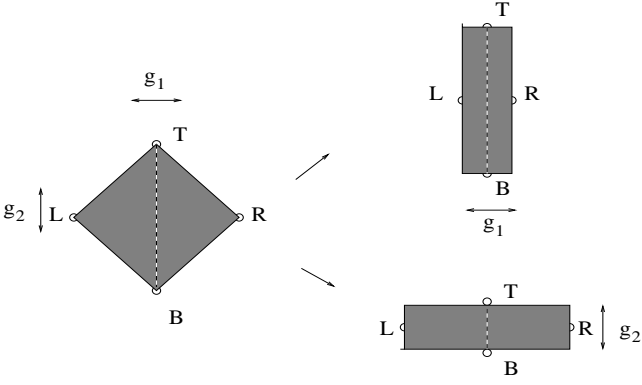}}
 \end{picture}
\end{center}
\caption{Schematic illustration of the split Hall bar model.  The edge of a quantum Hall system
with an interrupted line junction is mapped to a diamond.  The points T and B correspond to the
top and bottom of the opening in the line junction, the points between which top to bottom tunneling is
measured in the experiments. 
The interaction between two points that are the same distance from T (B) 
corresponds to interactions across the junction which we assume have relative strength $g_1$.
The points L and R correspond to the left and right ends of the Hall bar.  
The interaction between two points that are the same distance from L (R) 
corresponds to interactions across the Hall bar that we assume have relative strength $g_2$.
We measure position along the edge so that T is at $x=0$, R at $x=L/4$, B at $x=L/2$, and L at $x=3L/4$.
When (a) $g_1=g_2$ (diamond) the T to B quasiparticle tunneling scaling dimension is same as 
in the non-interacting 
 case, $d_{\rm tunn}=\nu$. At (b) $g_1\neq 0$, $g_2 = 0$, 
the scaling dimension of the tunneling operator between points T and B is 
$d_{\rm tunn}= \nu /K$, where $K$ defined in text.
 In (c) $g_1= 0$, $g_2 \neq 0$, and  
for the tunneling operator between points T and B we have
$d_{\rm tunn} = \nu K$. }
 \label{fig:two}
 \end{figure}
\bea
S&=& \frac{1}{L} \sum_{q_i} \int_{\omega} \phi(-q_i,-\omega)
\Bigl(\frac{i\omega}{\nu}{\rm sign}\; q_i + v_F |q_i| \Bigr)\phi(q_i,\omega)
\nonumber
\\[1mm]
\Bigl.
&& -\phi(q_i,-\omega)\left\{v_F |q_i| \left[g_1 + (-1)^{i} g_2\right]\right\} \phi(q_i,\omega)
\; .
\eea
It is convenient to introduce the parameters $\theta_\pm$, 
such that 
$\tanh2\theta_\pm= g_\pm$, where $g_\pm =g_1 \pm g_2$.  This action couples only 
wavevectors with equal or opposite momentum and may be diagonalized
analytically.  It yields two types of collective modes with resonant frequencies 
$\omega_{-}= \tilde{v}_{F_-} \nu |q_{2i-1}|$ at odd Fourier harmonics and
$\omega_{+}= \tilde{v}_{F_+} \nu |q_{2i}|$ at even Fourier harmonics, and 
with corresponding Fermi velocities $\tilde{v}_{F_\pm}=v_F/\cosh^22\theta_\pm= 
v_F \left[1-g_\pm^2\right]^{1/2}$.
As usual, the interactions renormalize the velocities downward.
Physically appropriate interaction parameters will normally satisfy the 
stability criterion $|g_\pm| < 1$.  Even and odd Fourier component fields have 
different correlations and, as we discuss below, enter the physically 
relevant correlations functions in different ways.  

\section{Correlation Functions of the Split Hall Bar Model}

The aim of this section is to derive expressions for the correlation 
functions of operators that describe tunneling of integer or fractional
charges between different points on the edge of the split Hall bar model.   
We start by evaluating some of the constituent components 
of the more complex correlation functions.  
Correlation functions can be evaluated using a path
integral formalism or an equivalent operator formalism that we find 
more convenient for these calculations.

\subsection{Phase Field Correlation Functions} 

We first show that in the presence of interactions the chiral edge fields can be expressed as a sum of 
components with opposite chirality, each of which splits into even and odd Fourier component pieces: 
$\phi(x,t) = \phi_-(z_-,\bar{z}_-) + \phi_+(z_+,\bar{z}_+)$.
Here $z_\pm = x-\tilde{v}_{F_\pm}t$ and $\bar{z}_\pm = x+\tilde{v}_{F_\pm}t$.
We start from the expression for $\phi(x,t)$ in terms of the elementary excitation creation 
and annihilation operators of the $g_1=g_2=0$ edge, 
$b_n=\phi(q_n)/\sqrt{L}$:
\bea
\phi(x) = \sum_{n>0}\sqrt{\frac{\nu}{ 4\pi n } } \left[b_n e^{i q_n x} + b^\dagger_n e^{-i q_n x}\right] 
\quad .
\eea
The $b_n$ are related to the full interaction model elementary excitation creation and
annihilation operators, $\chi(q_n)$ and $\chi^\dagger(q_n)$, by separate Bogoliubov transformations
for even and odd $q_n$.  For even $n$
\bea
b_{2n} &=& \cosh \theta_+ \chi(q_{2n}) + \sinh\theta_+ \chi^\dagger(q_{2n}) \quad, 
\eea  
and the corresponding contribution to the phase field is
\bea
\label{even_field}
\phi_+(x) & = &\sum_{n>0} \sqrt{\frac{\nu}{4\pi (2n)}} 
\\[0mm]\nonumber
 &&       
\times\left\{\left[e^{iq_{2n}x}\chi^\dagger(q_{2n}) + e^{-iq_{2n}x} \chi(q_{2n})\right]\sinh \theta_+  
\right.
\\[0mm] \nonumber
&&  \left.+ \left[e^{-iq_{2n}x}\chi^\dagger(q_{2n}) + e^{iq_{2n}x}\chi(q_{2n}) \right]
\cosh\theta_+  \right\}  \, .
\eea
Since the time dependence of the elementary excitation field 
can be taken into account by  the simple phase factor
$\chi(q_{2n})  \rightarrow  \chi(q_{2n})\exp\{-iq_{2n}v_{F_+}t\}$,
$\phi_+(x,t)$ can be written as a sum of fields with opposite chiralities
\bea
\phi_+(z_+,\bar{z}_+)=\cosh\theta_+ \, \chi_+(z_+) + \sinh\theta_+ \, \bar{\chi}_+(\bar{z}_+) \quad ,
\label{R_L_split}
\eea
where $\chi_+(z_+)$ and $\bar{\chi}_+(\bar{z}_+)$ can be read off from Eq.~(\ref{even_field}). 
In the limit of $g_1=g_2 \rightarrow 1/2$ the opposite chirality parts of $\phi_+(z_+,\bar{z}_+)$ 
have identical coefficients.
Similar remarks apply for the odd harmonic field which can be written as 
$\phi_-(z_-,\bar{z}_-)=\cosh\theta_- \, \chi_-(z_-) + \sinh\theta_- \, \bar{\chi}_-(\bar{z}_-) $.
The left moving components $\bar{\chi}_\pm(\bar{z}_\pm)$ vanish when interactions are such that $\theta_\pm\to 0$.

Physical observables can be expressed in terms of correlations of the phase field,
\bea
\left< \phi(x,t) \phi(x',t')\right> &=& \bigl<\left[\phi_-+\phi_+\right]_{(x,t)}
\left[\phi_-+\phi_+\right]_{(x',t')} \bigr> ,
\nonumber\\[3mm]
&=&
\sum_{\alpha=\pm}\left<\phi_\alpha(z_\alpha,\bar{z}_\alpha)\phi_\alpha(z_\alpha,\bar{z}_\alpha)\right> ,
\eea
with the second form for the right end side following from the independence of even and odd
harmonics.
For both even and odd fields, correlations between components with opposite chirality are 
finite and both contributions must be written as the sum of four terms.  For even $n$,
\bea
\label{temp_correl}
\left<\phi_+(z_+,\bar{z}_+) \phi_+(z_+',\bar{z}_+')\right> \hspace{-1.5mm} 
&=& \hspace{-1.5mm}\sum_{n>0} \frac{\nu}{4\pi (2n)} 
\left[e^{iq_{2n}(z_+-z'_+)}
\bar{N}_{2n} 
\right.
\nonumber \\[1mm]\nonumber
&+& \left.  \hspace{-1.5mm} e^{-iq_{2n}(z_+-z'_+)} N_{2n} \right] \cosh^2\theta_+
\\[1mm]\nonumber
&+&\hspace{-1.5mm}
\left[e^{-iq_{2n}(\bar{z}_+-\bar{z}'_+)} \bar{N}_{2n} 
\right.
\\[1mm]\nonumber
&+& \left.\hspace{-1.5mm}
  e^{iq_{2n}(\bar{z}_+-\bar{z}'_+)} N_{2n} \right] \sinh^2\theta_+
\\[1mm]\nonumber
&+& \hspace{-1.5mm}\left[e^{iq_{2n}(z_++\bar{z}'_+)} \bar{N}_{2n} 
\right.
\\[1mm]\nonumber
&+&\left.\hspace{-1.5mm}
  e^{-iq_{2n}(z_++\bar{z}'_+)} N_{2n} 
\right]\frac{1}{2} \sinh2\theta_+ 
\\[1mm]\nonumber
&+& \hspace{-1.5mm}\left[ e^{-iq_{2n}(\bar{z}_+ +z'_+)} \bar{N}_{2n} 
\right.
\\[1mm]
&+& \left. \hspace{-1.5mm}e^{iq_{2n}(\bar{z}_+ +z'_+)} N_{2n} 
\right]\frac{1}{2} \sinh2\theta_+ . 
\nonumber\\*
\eea
The final two terms are due to correlations between partial fields with opposite chirality.
Since they do not have translational invariance  
we refer to them in the following as {\em boundary} terms.
In Eq.~(\ref{temp_correl}) 
\bea
N_{2n} = \bigl<\chi^\dagger(q_{2n})\chi(q_{2n})\bigr> = \frac{1}{e^{(2\pi/L)\tilde{v}_{F_+}2n \beta}-1} \quad,
\label{Nq}
\eea 
is the elementary excitation Bose factor and $\bar{N}_{2n}  = \bigl<\chi(q_{2n})\chi^\dagger(q_{2n})\bigr> = 1 + N_{2n}$.
The correlation function of odd Fourier component fields, $\left<\phi_-(x,t)\phi_-(x',t')\right>$, is obtained from the above by substitution of
$z_-$, $\bar{z}_-$ for $z_+$ and $\bar{z}_+$, and $\theta_-$ for $\theta_+$.

All the sums over Fourier components above have the same form.  We define 
\begin{eqnarray}
\label{phase_sum}
S_{\rm e/o}(\Delta z) &=& \sum_{n>0}^{\rm e/o} \frac{1}{4\pi n}
\left[\left(e^{iq_{n} \Delta z} -1\right) \bar{N}_{n} 
\right.
\nonumber \\
&&\left.+ \left(e^{-iq_{n}\Delta z} -1\right)N_{n} \right]
,
\end{eqnarray}
where $q_n=nq_1$ and $q_1=2\pi/L$.  Technical details of the evaluation
of this sum are explained in the Appendix.  We find that   
\bea
\label{even_sum}
S_{\rm e}(\Delta z_+) = &-& \frac{1}{8\pi}\ln\left[
       {\rm e}^{{\rm i}(\frac{2\pi\Delta z_+}{L}-\frac{\pi}{2})}
\frac{ \vartheta_1\left(\frac{\pi \Delta z_+}{L}|  i\frac{v\beta}{L}\right)  }
     { \vartheta_1\left(\frac{\pi \alpha}{L}|i\frac{v\beta}{L} \right)   }   
\right.
\nonumber \\[3mm]
&\times & \left. \frac{ \vartheta_2\left(\frac{\pi \Delta z_+}{L}|  i\frac{v\beta}{L}\right)  }
     { \vartheta_2\left(\frac{\pi \alpha}{L}|i\frac{v\beta}{L} \right)   }   
\right],
\eea
and 
\begin{equation}
\label{odd_sum}
S_{\rm o}(\Delta z_-) \hspace{-1mm} = \hspace{-1mm} - \frac{1}{8\pi}\ln\left[
       {\rm e}^{-{\rm i}\frac{\pi}{2}}
\frac{ \vartheta_1\left(\frac{\pi \Delta z_-}{L}|  i\frac{v\beta}{L}\right)  }
     { \vartheta_1\left(\frac{\pi \alpha}{L}|i\frac{v\beta}{L} \right)   }
\frac{ \vartheta_2\left(\frac{\pi \alpha}{L}|i\frac{v\beta}{L} \right)   } 
     { \vartheta_2\left(\frac{\pi \Delta z_-}{L}|  i\frac{v\beta}{L}\right)  }
\right]
\end{equation}
where $\vartheta_1$ and $\vartheta_2$ are the elliptic theta functions\cite{gradshteyn,stegun}.

In the following equations we define
$F(\Delta z_\pm)=\vartheta_1({\pi \Delta z_\pm/L}|  i{v\beta/L})/\vartheta_1({\pi \alpha/L}|i{v\beta/L} )$ and 
$\tilde{F}(\Delta z_\pm) = 
\vartheta_2({\pi \Delta z_\pm/L}|  i{v\beta/L}) / \vartheta_2({\pi \alpha/L}|i{v\beta/L} )$.
Below we will use the property that $F(\Delta z_\pm)\sim \Delta z_\pm$ for $\Delta z_\pm \ll L$ whereas 
$\tilde{F}(\Delta z_\pm) \sim 1$ in that limit.
With this notation, the argument of the logarithm in the even-component summation Eq.~(\ref{even_sum}) is given up to a phase 
factor by the product $F(\Delta z_+)\tilde{F}(\Delta z_+)$ whereas the argument of the logarithm of the odd-component summation
equals $F(\Delta z_-)/\tilde{F}(\Delta z_-)$ up to a phase. 
It therefore follows from Eq.~(\ref{even_sum}) and Eq.~(\ref{odd_sum}) that
adding even and odd components yields 
sums of $\ln F(\Delta z_+)$ and $\ln F(\Delta z_-)$ and differences of $\ln \tilde{F}(\Delta z_+)$ and 
$\ln \tilde{F}(\Delta z_-)$.

For the terms containing 
$\sinh\theta_\pm \cosh\theta_\pm$ [fourth and fifth line of Eq.~(\ref{temp_correl})], the result can be
obtained using also Eqs.~(\ref{even_sum}) and (\ref{odd_sum}).
In both of these terms space coordinates appear in the combinations $z_\pm+\bar{z}_\pm$, $z'_\pm+\bar{z}'_\pm$, 
and $z_\pm+\bar{z}'_\pm$, $\bar{z}_\pm+z'_\pm$, which are not translational invariant.  

\subsection{One-Particle Green's Functions} 

The one-particle Green's function expressions follow from the phase field correlation
functions by applying the identity 
stated in the Appendix;  [see Eq.(\ref{campbell})]. We find that 
\begin{widetext}
\bea
\label{greens_funct}
\bigl<R_{\rm QP}^\dagger(x,t) R_{\rm QP}(x',t')\bigr> &=&
\left[F^{-\frac{\nu}{2}\cosh^2\theta_+}(z_+ - z_+')F^{-\frac{\nu}{2}\cosh^2\theta_-}(z_- - z_-')\right]
\left[F^{-\frac{\nu}{2}\sinh^2\theta_+}(\bar{z}_+ - \bar{z}_+')F^{-\frac{\nu}{2}\sinh^2\theta_-}(\bar{z}_- - \bar{z}_-')\right]
\nonumber\\[3mm]\nonumber
& & \hspace{-25mm} \times \left[\frac{\tilde{F}^{-\frac{\nu}{2}\cosh^2\theta_+}(z_+ - z_+') 
              \tilde{F}^{-\frac{\nu}{2}\sinh^2\theta_+}(\bar{z}_+ - \bar{z}_+')
}{            \tilde{F}^{-\frac{\nu}{2}\cosh^2\theta_-}(z_- - z_-')
              \tilde{F}^{-\frac{\nu}{2}\sinh^2\theta_-}(\bar{z}_- - \bar{z}_-')
}\right]
\left[ \frac{F^{-\frac{\nu}{2}\sinh\theta_+ \cosh\theta_+}(z_++\bar{z}_+')
             F^{-\frac{\nu}{2}\sinh\theta_- \cosh\theta_-}(z_-+\bar{z}_-')}
{            F^{-\frac{\nu}{2}\sinh\theta_+ \cosh\theta_+}(z_++\bar{z}_+) 
             F^{-\frac{\nu}{2}\sinh\theta_- \cosh\theta_-}(z_-+\bar{z}_-)  }\right]
\\[3mm]\nonumber
&& \hspace{-25mm} \times
\left[ \frac{F^{-\frac{\nu}{2}\sinh\theta_+ \cosh\theta_+}(\bar{z}_+ + z_+')
             F^{-\frac{\nu}{2}\sinh\theta_- \cosh\theta_-}(\bar{z}_- +z_-')}
{            F^{-\frac{\nu}{2}\sinh\theta_+ \cosh\theta_+}(\bar{z}_+'+ z_+') 
             F^{-\frac{\nu}{2}\sinh\theta_- \cosh\theta_-}(\bar{z}_-'+ z_-')  }\right]
\\[3mm]
&&  \hspace{-25mm} \times
\left[\frac{\tilde{F}^{-\frac{\nu}{2}\sinh\theta_+ \cosh\theta_+}(z_+       +  \bar{z}_+')
      \tilde{F}^{-\frac{\nu}{2}\sinh\theta_+ \cosh\theta_+}(\bar{z}_+ +  z_+')
}{    \tilde{F}^{-\frac{\nu}{2}\sinh\theta_- \cosh\theta_-}(z_-+\bar{z}_-')
      \tilde{F}^{-\frac{\nu}{2}\sinh\theta_- \cosh\theta_-}(\bar{z}_- + z_-')}
\right]
\nonumber
\\[3mm]
&& \hspace{-25mm} \times
\left[
\frac{\tilde{F}^{-\frac{\nu}{2}\sinh\theta_- \cosh\theta_-}(z_-  +  \bar{z}_-)
      \tilde{F}^{-\frac{\nu}{2}\sinh\theta_- \cosh\theta_-}(z_-' +  \bar{z}_-')
}{    \tilde{F}^{-\frac{\nu}{2}\sinh\theta_+ \cosh\theta_+}(z_+  +  \bar{z}_+)
      \tilde{F}^{-\frac{\nu}{2}\sinh\theta_+ \cosh\theta_+}(z_+' +  \bar{z}_+')}
\right] .
\eea
\end{widetext}
 It is worth pointing out that as ($g_-\rightarrow 0$, $g_+\rightarrow 1$)
 $\theta_- \rightarrow 0$ and $\theta_+ \rightarrow +\infty$, the fast modes
 exhibit translational symmetry and decay with the non-interacting exponent.
 For the slow moving modes translational invariance is strongly broken
 and the exponents are far from the values of the non-interacting case.

The correlation functions depend on the proximity to the points $L$, $R$, $T$, and $B$
in Fig.~2.  It is instructive to consider explicitly the special cases in which one or the other of
the two interaction parameters vanishes:
\noindent 
(a) $g_1 \neq 0$, $g_2=0$, or $\theta_-=\theta_+ = \theta$, and
\bea
\bigl<R_{\rm QP}^\dagger(x,t) R_{\rm QP}(x',t')\bigr> \hspace{-2mm} & = & \hspace{-2mm}
\left[ \frac{F(z+\bar{z}')F(\bar{z}+z' )}
{            F(z+\bar{z}) F(\bar{z}'+z')} \right]^{-({\nu}/{2})\sinh 2\theta }
\nonumber\\[2mm]
&& \hspace{-20mm} \times
 \left[F^{-\frac{}{}\nu\cosh^2\theta}(z - z')F^{-\frac{}{}\nu\sinh^2\theta}(\bar{z} - \bar{z}')\right]
.
\nonumber\\*
\label{F}
\eea
\noindent 
(b) $g_1=0$, $g_2\neq 0$, or $\theta_-=-\theta_+ = -\theta$, and
\bea
\bigl<R_{\rm QP}^\dagger(x,t) R_{\rm QP}(x',t')\bigr> \hspace{-2mm} & = & \hspace{-2mm}
\left[ \frac{\tilde{F}(z+\bar{z}')\tilde{F}(\bar{z}+z' )}
{            \tilde{F}(z+\bar{z}) \tilde{F}(\bar{z}'+z')} \right]^{-(\nu/2)\sinh 2\theta }
\nonumber
\\[3mm]
&& \hspace{-20mm} \times
\left[F^{-\frac{}{}\nu\cosh^2\theta}(z - z')F^{-\nu\frac{}{}\sinh^2\theta}(\bar{z} - \bar{z}')\right]
\;.
\nonumber\\*
\label{F_tilde}
\eea

Important properties of the Green's function follow from properties of the constituent elliptic
$\vartheta$ functions. Most importantly $ \vartheta_i(u\pm \pi) = \mp \vartheta_i(u)$, where $1=1,2$,
and $\vartheta_1(u\pm {\pi}/{2}) = \pm \vartheta_2(u)$.
Inspection of these equations verifies that the quasiparticle correlation function exhibit
the correct periodicities in space, time and temperature;
periodicity $L$ in space $\bigl<R^\dagger(mL+z) R(z')\bigr>=\bigl<R^\dagger(z) R(z')\bigr>$,
for integer $m$, quasi-periodicity in time $T_\pm=L/v_\pm$ and
quasi-periodicity as a function of temperature with ${\cal T}_\pm = i \pi v_\pm \beta/L$.
Notice also the property that the correlation functions transform to each other if we shift 
$z,z'\rightarrow z+L/4,z'+L/4$.
If $x$ and $x'$ fall in the boundaries of (a), 
they will fall in the bulk of (b) and would go back in the boundaries under another shift of the coordinates by 
$L/4$.

We discuss one example of how these exponents are different from the case of the usual 
Luttinger liquids. 
Consider the case $g_1-g_2=0$ and $g_1+g_2 \neq 0$. In this case we have 
$\theta_-=0$ and expect $\theta_+ \gg 1$. The scaling dimension of the static fermion-fermion correlation 
function for $0 \sim x' \ll x \ll L$ is 
\begin{equation}
d = \frac{\nu}{2}\left[1+\cosh2\theta_+ + \frac{1}{2}\sinh2\theta_+ \right] 
  = \frac{\nu}{8}\left[4 + K + \frac{3}{K}\right]  \,,
\end{equation}
where $K=\exp\{-2\theta_+\}$. At strong interaction strengths, in this case,
the quasiparticle-quasiparticle correlation functions decay slower then in the usual LL case
for which $d=(\nu/2)(K+1/K)$. 

\subsection{Correlation functions of the quasiparticle tunneling operator}

So far we have been discussing the case of an open constriction in which the Hall liquid is extended from
left to right through the constriction. In this case in addition to electrons, quasiparticles can 
also backscatter to the opposite chirality edge since they can live within the QH liquid. 
The objective of the following calculation is to determine the relevance of quasiparticle 
tunneling processes between $T$ and $B$. 
We find here the correlation function between two tunneling operators 
$\hat{T}^{\rm QP}_{\rm tunn}(x,x';t)=R_{\rm QP}^\dagger(x,t)R_{\rm QP}(x',t)$.
For the correlator $G^{\rm QP}_{\rm tunn}(t-t')  = \bigl<\hat{T}^{\rm QP}_{\rm tunn}({L/2},0;t) 
\hat{T}^{\rm QP}_{\rm tunn}(0,{L/2};t')\bigr>$ 
[see Eq.~(\ref{tunn_correl3}) in the Appendix] we obtain the following
\bea
\label{tunn_correl}
 G^{\rm QP}_{\rm tunn}(t-t') &=& 
 \frac{G(\frac{L}{2},0;t,t) G(\frac{L}{2},\frac{L}{2};t,t')}{G(\frac{L}{2},0;t,t')}
\nonumber\\[3mm]
&& \times
\frac{G(0,0;t,t') G(0,\frac{L}{2};t',t')}{G(0,\frac{L}{2};t,t')}
\quad ,
\eea
where $G(x,x';t,t') = \bigl<R_{\rm QP}^\dagger(x,t) R_{\rm QP}(x',t')\bigr>$.

In the following, we consider the general case of $g_1\neq 0$, $g_2\neq 0$, 
with the aim of summarizing results for the scaling dimensions of the tunneling operator. 
These calculations are simplified by using the symmetry related property that $F$ transforms to ${\tilde F}$ and vice versa 
when 
their argument is shifted by half the system size.
In the fermionic Green's function Eq.~(\ref{greens_funct}) both the {\em boundary} and the {\em bulk} 
factors are themselves the product of a factor of the form $F^{\gamma_+}(+)F^{\gamma_-}(-)$ and a
second factor of the form $\tilde{F}^{\gamma_+}(+)/\tilde{F}^{\gamma_-}(-)$, where the $+$ and $-$ superscripts
refer to even and odd $n$ contributions. 
As can be seen from Eq.~(\ref{tunn_correl}) for the correlation function between two tunneling operators, we need two types of
constituent fermionic Green's functions; first, $G(L/2,L/2;t,t')$, $G(\epsilon,\epsilon;t,t')$ and second 
$G(L/2,0;t,t')$, $G(0,L/2;t,t')$. 
For the first type, the exponents are obtained from the $F$ functions of Eq.~(\ref{greens_funct}).
For the second type of correlations the arguments of all the time dependent $F$ and $\tilde{F}$ 
functions will be $(L/2 + \epsilon,\epsilon)$ 
or $(\epsilon,L/2 + \epsilon)$. Therefore in this case all the $F$ functions become $\tilde{F}$ and vice-versa. As a result the correlation function, 
will now be given in terms of a product of $F^{\gamma_+}(+)/F^{\gamma_-}(-)$.  Therefore the scaling dimensions
of the tunneling process will depend on the difference of interaction strengths $g_1-g_2$ only as is shown below. 
The final result for the constituent Green's functions is as follows:
\begin{widetext}
\bea
\label{correl_tun1}
G(\frac{L}{2},0;t,t') = G(0,\frac{L}{2};t,t') 
\nonumber 
& = &  f(\tilde{F}(+)\tilde{F}(-))\cdot
\left[
\frac{        {F}^{-\frac{\nu}{2}\cosh^2\theta_+}(-v_+(t - t'))   }
{             {F}^{-\frac{\nu}{2}\cosh^2\theta_-}(-v_-(t - t'))   } \;
\frac{        {F}^{-\frac{\nu}{2}\sinh^2\theta_+}(v_+(t - t'))    }
{             {F}^{-\frac{\nu}{2}\sinh^2\theta_-}(v_-(t - t'))    }
\right]
\\[3mm]
& \times  &
\left[
\frac{        {F}^{-\frac{\nu}{4}\sinh2\theta_+ }(-v_+(t - t'))     }
{             {F}^{-\frac{\nu}{4}\sinh2\theta_- }(-v_-(t-t'))       } \;
\frac{        {F}^{-\frac{\nu}{4}\sinh2\theta_+ }(v_+(t - t'))      }
{             {F}^{-\frac{\nu}{4}\sinh2\theta_- }(v_-(t-t') )       }
\right]
\, .
\eea
Or asymptotically for $v_\pm(t-t')/L \ll 1 $
\bea
G(\frac{L}{2},0;t,t') &=& 
 \left[v_-^{\frac{\nu}{2} e^{2\theta_+}} 
v_+^{-\frac{\nu}{2} e^{2\theta_-}} i(t-t')\right]^{-\frac{\nu}{2}\left(e^{2\theta_+}-e^{2\theta_-}\right)} 
\hspace{-7mm} .
\eea
Similarly
\bea
\label{correl_tun2}
G(\frac{L}{2},\frac{L}{2};t,t') &=& G(\epsilon,\epsilon;t,t')  
\nonumber\\[3mm]\nonumber
&=&
                                   f(\tilde{F}(+)/\tilde{F}(-))\cdot 
 \left[
            F^{-\frac{\nu}{2}\cosh^2\theta_+}(v_+(t' - t))    \;
            F^{-\frac{\nu}{2}\cosh^2\theta_-}(v_-(t' - t))
\right]
\\[3mm]
\nonumber
& \times &  \left[
            F^{-\frac{\nu}{2}\sinh^2\theta_+}(v_+(t - t'))    \;
            F^{-\frac{\nu}{2}\sinh^2\theta_-}(v_-(t - t'))
\right]
\left[ 
\frac{      F^{-\frac{\nu}{4}\sinh 2\theta_+ }(v_+(t'-t))   }
{           F^{-\frac{\nu}{4}\sinh 2\theta_+ }(2\epsilon)   } \;
\frac{      F^{-\frac{\nu}{4}\sinh 2\theta_- }(v_-(t'-t))   }
{           F^{-\frac{\nu}{4}\sinh 2\theta_- }(2\epsilon)   }
\right]
\\[3mm] 
&\times &
\left[ 
\frac{      F^{-\frac{\nu}{4}\sinh 2\theta_+ }(v_+(t-t'))  }
{           F^{-\frac{\nu}{4}\sinh 2\theta_+ }(2\epsilon)  } \;
\frac{      F^{-\frac{\nu}{4}\sinh 2\theta_- }(v_-(t-t'))  }
{           F^{-\frac{\nu}{4}\sinh 2\theta_- }(2\epsilon)  }
\right] \, .
\eea
\end{widetext}
In the asymptotic limit of $v_\pm (t-t')/L \ll 1 $
\bea
G(\frac{L}{2},\frac{L}{2};t,t') &=& G(\epsilon,\epsilon;t,t') 
\\[3mm]\nonumber
&=&
\left[v_-^{ \frac{\nu}{2} e^{2\theta_+}}
       v_+^{\frac{\nu}{2} e^{2\theta_-}} i(t-t')\right]^{-\frac{\nu}{2}\left(e^{2\theta_+} + 
                                                                         e^{2\theta_-}\right)}
\;.
\eea

Therefore the correlation function of two tunneling operators as a function of $z_-=v_-(t-t')$ 
will have the following form
 \bea
 \label{general_tunneling}
 G_{\rm tunn}^{\rm QP} (t-t') = 
                [\tilde{F}(z_-)/F(z_-)]^{2 \nu e^{2\theta_-}}
\hspace{-2mm} =
 {\left(iz_-\right)^{-2\nu e^{2\theta_-}}} 
\hspace{-1.5mm}\,.
 \eea

The tunneling scaling dimension is given by 
\bea
\label{tunn}
d = \nu e^{2 \theta_-} \quad,
\eea
where $\tanh 2\theta_- = g_1 - g_2 $. 
We see that in the experimental setup of Fig.~1 this exponent is no longer universal. 
Notice also  the duality
$d(g_1,g_2) = \nu^2/ d(g_2,g_1)$,
from which it follows that $d(g_1,g_1) = \nu $.
This duality under the exchange $g_1$ and $g_2$ is related to the duality between angles $\alpha$ and $\pi-\alpha$ 
in $X$-shaped constriction model studied by Pryadko, Shimshoni and Auerbach\cite{pryadko}.

It is quite surprising that in the case $g_2=g_1$, the tunneling exponent is the one corresponding to 
the noninteracting case. We also notice that although the 
fermion-fermion correlation function decays very quickly in this limit, the charge-charge density correlation function 
decays with an exponent that equals that of the non-interacting case. 
The quasiparticle charge-charge density correlation functions decay with same exponent:
\bea
\left< \rho(t) \rho(t') \right> = \frac{1}{\pi}\frac{1}{[iv_-(t-t')]^{2\nu}} \quad .
\eea
The same happens for the tunneling-tunneling correlation function.

From 
Eq.~(\ref{tunn}) we get the following special cases that we have used before
(a) $g_1\neq 0$, $g_2 = 0$, with $\theta_-=\theta_+$. Substituting this in  Eq.~(\ref{general_tunneling}) we get
for the scaling dimension of the tunneling operator close to the origin
$d_{\rm tunn}^{\rm bnd.}= \nu /K$,
where $K=e^{-2\theta_+}$.   
(b) $g_1= 0$, $g_2 \neq 0$, with $\theta_-=-\theta_+$. 
For the tunneling operator close to the origin we get, 
$d^{\rm bulk}_{\rm tunn} = \nu K$,
where $K$ is given by same formula. 

The same results can also be obtained by using the path integral approach. Here one makes use of matrices $M$ and 
$M^\dagger$ that diagonalize the action and fulfill $MJM^\dagger=J$, where $J=\delta_{ij}{\rm sign} j$. These 
matrices have elements along the two main diagonals (thus breaking translation invariance)
with the even and odd components depending on parameters 
$\theta_+$ and $\theta_-$, respectively. In tunneling between points $x=0$ and $L/2$ only the 
odd components contribute in tunneling correlation functions.

\subsection{Tunneling conductance}

In this section we calculate the quasiparticle inter-edge tunneling and the differential tunneling conductance 
for this model. 
This can be done perturbatively for small tunneling amplitudes by using Fermi's golden rule. 
The following derivation is slightly different from the one that Kane and Fisher\cite{kane} (KF in the 
following) have used.
The differences arise because, unlike the KF case, our chiral left-right movers have a
finite average value for  their correlations and there are six Green's functions, Eq.~(\ref{tunn_correl}), instead of the two 
of KF.  Our results agree when we switch off one of the interactions however.
The quasiparticle current across the Hall bar at the constriction in the presence of 
a top-to-bottom source-drain bias voltage can be calculated starting from the Golden-rule expression, 
\begin{equation}
I^{\rm QP}_{\rm tunn} 
=\frac{2\pi e^*}{\hbar} \sum_n s_n \left|\left<n\right|H_{\rm tunn}\left|0\right>\right|^2 
\delta\left(E_n -E_0 -s_n eV\right) \, .
\end{equation}
The $n$-summation extends over many-body states in which an electron has been transferred across the 
constriction in the $s_n=\pm1$ direction\cite{kane}. 
In our case the tunneling operator is composed of
\begin{equation}
\label{H_tunn}
\hat{H}_{\rm tunn}  =   
{\cal T} R^\dagger_{\rm QP}(\frac{L}{2},0)R_{\rm QP}(0,0) +  {\cal T}^* R_{\rm QP}^\dagger(0,0)R_{\rm QP}(\frac{L}{2},0) \, ,
\end{equation}
or $\hat{H}_{\rm tunn} = \hat{T}_{\rm tunn}^{\rm QP}({L/2},0;0) + {\rm H.c}$.
In Eq.~(\ref{H_tunn}) ${\cal T}$ is the tunneling amplitude.
We find that 
\begin{equation}
I^{\rm QP}_{\rm tunn} = \left[1-\exp\left(-\frac{eV}{k_BT}\right)\right]  \frac{e^*}{\hbar}
\int_{-\infty}^{+\infty} \rd t e ^{ieVt} \;
G^{\rm QP}_{\rm tunn}(t) \,,
\label{I_tunn}
\end{equation}
where $G^{\rm QP}_{\rm tunn}(t)$ was evaluated in the previous section [see Eq.~(\ref{general_tunneling})]. 
The large system size and finite temperature tunneling correlation function is
\be
G^{\rm QP}_{\rm tunn}(t-t') = \frac{|{\cal T}|^2}{(2\pi)^2} \left( i \frac{v_- \beta}{\pi}  
\sinh \frac{\pi (t-t')}{\beta}      \right)^{-2\nu e^{2\theta_-}} 
\hspace{-3mm},
\ee
which reduces at zero temperature to
\bea
G^{\rm QP}_{\rm tunn}(t-t') = \frac{|{\cal T}|^2}{(2\pi)^2} \Bigl[ i v_- (t-t') \Bigr]^{-2\nu e^{2\theta_-}} 
 .
\eea
After taking the Fourier transform of these correlation functions for the large $L$ and finite temperature limit for 
the $s=\pm$ currents we obtain\cite{beta}
\bea
\label{Iplus}
I^{\rm QP}_\pm &=& 
 \frac{e^*}{\hbar} \frac{|{\cal T}|^2}{(2\pi)^2} \frac{1}{v} \left(\frac{v \beta}{2\pi}\right)^{-2d}
\nonumber\\[2mm]
&&\times 2{\rm Re}\left[ (-i)^{2d} B\left(d \mp i\frac{eV\beta}{2\pi},1-2d\right)\right],
\eea
with the total tunneling current being
\bea
I^{\rm QP}_{\rm tunn} &=&
 \frac{ e^* \sin(\pi d) }{\hbar} \frac{|{\cal T}|^2}{(2\pi)^2}
\frac{1}{v} \left(\frac{v \beta}{2\pi}\right)^{-2d}
\nonumber \\[3mm] 
&& \times 4 {\rm Im} \left[B\left(d+i\frac{eV\beta}{2\pi},1-2d\right)\right]\quad .
\label{I_tot}
\eea
In the above equations $B(\cdot)$ is the Euler beta function. 

For the zero temperature case the chiral currents reduce to
\bea
I^{\rm QP}_{\pm} = \frac{ e^*}{\hbar} \int_{-\infty}^{+\infty} \rd t 
\frac{|{\cal T}|^2}{(2\pi)^2 v_-^{2\nu e^{2\theta_-}} }
\frac{e^{\pm i eVt}}{[i(t-i\alpha)]^{2\nu e^{\theta_-}} } \quad ,
\eea
where in the above we have put back the regularization constant $-i\alpha$. 
Notice that the interval of integration is doubled and the real part is taken out 
(the integral is real). 
This integration is done by introducing a branch cut in the complex $t$-plane
in the positive imaginary axis starting from the point $i \alpha$ going to infinity.
Contributions will be obtained by closing the contour on the upper half plane for $V>0$ in 
case of $I^{\rm QP}_+$ and for $V<0$ in case of $I^{\rm QP}_-$. $I^{\rm QP}_\pm$ are given by
\bea
I^{\rm QP}_{\pm} &=& \Theta(\pm V) \frac{ e^* |{\cal T}|^2}{h v_-^{2\nu e^{\theta_-}} }
\frac{(\pm eV)^{2\nu e^{\theta_-}-1} }{\Gamma(2\nu e^{\theta_-})  }
\quad .
\eea

The behavior of the tunneling current and tunneling conductance for $eV\beta/2\pi \ll 1$ can be obtained by using
\cite{low_V} to give
\begin{equation}
I^{\rm QP}_{\rm tunn}  = C
\frac{eV\beta}{2\pi}\left[1+\left(\frac{\pi^2}{6}-\zeta(2,d)\right)
\left(\frac{eV\beta}{2\pi}\right)^2 +\cdots\right]
\quad,
\end{equation}
where $\zeta\left(2,d\right)$ is the Riemann
zeta function and
\bea
C = \frac{ e^* |{\cal T}|^2}{v_- h} \left(\frac{v_-\beta}{2\pi} \right )^{1-2d}
\frac{\Gamma^2(d)}{\Gamma(2d)} \quad .
\eea

The significance of $d$ in the quasiparticle tunneling current expression is more apparent
in the simpler expressions for low temperature (high bias voltage) and high temperature (low bias voltage)
limits.  For $eV\beta/2\pi \ll 1$
the non-linear part of the conductance is positive for $d>1$ and negative for $d<1$.  The latter case is the one
studied by KF and at $V\ll T$ the top-bottom linear tunneling conductance diverges as $T^{2d-2}$,
with the leading non-linear correction negative and proportional to 
$T^{2d-4}V^2$. In the opposite case when $d>1$ the top-bottom linear 
tunneling conductance goes to zero as $T^{2d-2}$ while the leading non-linear correction 
is positive and varies as $T^{2d-4}V^2$.

For finite system sizes and finite temperatures the quasiparticle tunneling correlation function takes the form 
\bea
G^{\rm QP}_{\rm tunn}(t)= \frac{|{\cal T}|^2}{(2\pi)^2} F^{-2\nu e^{2\theta_-}}(z_-)
\tilde{F}^{2\nu e^{2\theta_-}}(z_-) \;,
\label{I_Tunn_sum2}
\eea 
which at zero temperature reduces to
\bea
G_{\rm tunn}^{\rm QP}(t) =
\frac{ |{\cal T}|^2 }{ {(2\pi)^2} } 
\left[i\frac{L}{2\pi}\sin\left(\frac{2\pi v_- t}{L}\right)\right]^{-2\nu e^{2\theta_-}}
\quad .
\eea
See the Appendix for a discussion of the corresponding $I$-$V$ expression.  
At finite sizes the current is composed of $\delta$-function peaks with coefficients that are given in 
the Appendix. 
This reflects the fact that at finite sizes the momenta and energy levels are discrete. 
The tunneling current for this case is given by
\bea
I^{\rm QP}_+  &=&\frac{ |{\cal T}|^2 }{ {(2\pi)} } \left(\frac{L}{2\pi}\right)^{-2d}
\nonumber\\[2mm]
&\times & \hspace{-2mm }\sum_{n=0}^\infty {{2d +n-1}\choose {n}}
\delta\left[eV - \frac{4\pi v_- }{L}(n+d)\right] ,
\eea
where for large $L$ ($n/L$ constant) 
\bea
{{2d +n-1}\choose {n}} \sim n^{2d-1}.
\nonumber
\eea

\subsection{Closed constriction limit}

In the following we look briefly at the case of the closed constriction limit.  In this limit 
we can easily allow for different filling fractions and Fermi velocities in the decoupled 
quantum Hall effect systems.  In this limit we should consider only electron tunneling, since quasiparticles 
cannot tunnel through the vacuum.
The electron tunneling operator in this case is $\hat{T}^{\rm el}_{\rm tunn}(x=0,t)=R^\dagger(0,t) L(0,t)+ {\rm H.c.}$, 
where
\be
L(x)=\frac{1}{\sqrt{2\pi}} e^{ i\frac{\sqrt{4\pi}}{\nu_L} \phi_L(x)} \; , \;
R(x)=\frac{1}{\sqrt{2\pi}} e^{-i\frac{\sqrt{4\pi}}{\nu_R} \phi_R(x)} \; .
\label{electron_ops}
\ee
Again from the identity Eq.~(\ref{campbell}) for the left-right electron tunneling correlation function we have
\begin{equation}
\label{tunn_correl2}
G^{\rm el}_{\rm tunn}(t-t') = \hspace{-1mm}
\frac{G_{R^\dagger L^\dagger}(t,t') G_{R^\dagger R}(t,t')}{G_{R^\dagger L}(t,t)}\frac{G_{LL^\dagger }(t,t') 
G_{LR}(t,t')}{G_{L^\dagger R}(t',t')} \, ,
\end{equation}
where $G$ is the usual single fermion Green's function
$G_{RR}(x,x';t,t') = \bigl<R^\dagger(x,t) R(x',t')\bigr>$,
all taken at the point of tunneling, $x=0$.
These Green's functions can be evaluated as in previous sections to give
\begin{eqnarray}
\label{el_tunn_CF}
G_{R^\dagger R}(t) 
&=& F^{- \frac{\sinh^2\theta}{\nu_L}}(\bar{z}_L-\bar{z}'_L)  F^{-\frac{\cosh^2\theta}{\nu_R} }(z_R-z'_R) \, ,
\nonumber \\[3mm]
G_{L^\dagger L}(t) &=& F^{- \frac{\cosh^2\theta}{\nu_L} }(\bar{z}_L-\bar{z}'_L)  
F^{- \frac{\sinh^2\theta}{\nu_R} }(z_R-z'_R) \, ,
\nonumber\\[3mm] \nonumber
G_{RL}(t) &=& F^{-\frac{\sinh2\theta}{2\sqrt{\nu_L\nu_R} } }(\bar{z}_L-\bar{z}'_L) 
F^{-\frac{\sinh 2\theta}{2 \sqrt{\nu_L\nu_R}} }(z_R-z'_R) \, ,
\\*
\end{eqnarray}
where $\bar{z}_L= x+v_Lt$, $z_R= x-v_Rt$.
In the above formulas the correlation functions for electrons on the edges of the left or the right 
QH liquids is composed of a product where one can recognize that one of the factors is the contribution from the 
opposite chirality component that appears in presence of interactions. Such terms are the ones containing the 
exponents $\sinh^2\theta$, $\sinh2\theta$, which reduce to unity when the interedge interactions vanish
$V_{\rm RL}\to 0$.

Using the above for electron tunneling between points $x=0$, $x'=0$,  we get
\bea
G^{\rm el}_{\rm tunn}(t)  &=&
\frac{|{\cal T}|^2}{(2\pi)^2}
F^{-\frac{1}{ \nu_L }\left(\cosh 2\theta+\sqrt{\frac{\nu_L}{\nu_R} }\sinh 2\theta   \right) } ( \bar{z}_L-\bar{z}'_L)
\nonumber\\[3mm] &&
\times F^{-\frac{1}{ \nu_R }\left(\cosh 2\theta+\sqrt{\frac{\nu_R}{\nu_L} }\sinh 2\theta   \right) } ( z_R-z'_R) \,.
\nonumber\\*
\eea
For large size systems we find that 
the {\em electron} tunneling-tunneling correlation function is
\be
G^{\rm el}_{\rm tunn} (t-t') = \frac{|{\cal T}|^2}{(2\pi)^2}
\prod_{\alpha=\pm} \left(i \frac{v_\alpha\beta}{\pi}\sinh\frac{\pi (t-t')}{\beta}\right)^{\eta_\alpha} 
\hspace{-2mm} ,
\ee
where $\alpha = \pm$ for $R$, $L$, respectively, and
\bea
\eta_\pm= -\frac{1}{\nu_\pm} \left[\cosh 2 \theta- \left(\frac{\nu_\pm}{\nu_{\mp}}\right)^{1/2}\sinh 2\theta\right]
\quad . 
\eea
When $\nu_L=\nu_R$, the result for the scaling dimension will be $d^{\rm el} = ({1}/{\nu}) e^{2\theta}$,
where $\theta$ is defined by $\tanh 2\theta = -2V_{RL}/[V_{LL}+V_{RR}]$. $V_{\rm RR}$ and $V_{\rm LL}$ are 
intra-edge interactions on the right and left QH liquids, respectively. Since the $\theta$ defined by 
this equation is negative, the scaling dimensions of this process is inverse to the one defined 
in the singly connected loop of the previous sections. The top-bottom QP
tunneling process for an open constriction is relevant whenever left-right
{\em electron} tunneling through a closed constriction is irrelevant and vice-versa. 

Nevertheless this result is another equivalent way of expressing the duality observed in 
Eq.~(\ref{tunn}) 
under the exchange $g_1 \leftrightarrow g_2$ and $\nu \leftrightarrow 1/\nu$.
The latter expresses the quasiparticle-electron exchange that necessarily 
takes place when the constriction is closed.

\section{Discussion}

\begin{figure}
\unitlength=1mm
\begin{center}
\begin{picture}(70,52)
\put(-1,50){\includegraphics[height=58\unitlength,width=73\unitlength,angle=-90]{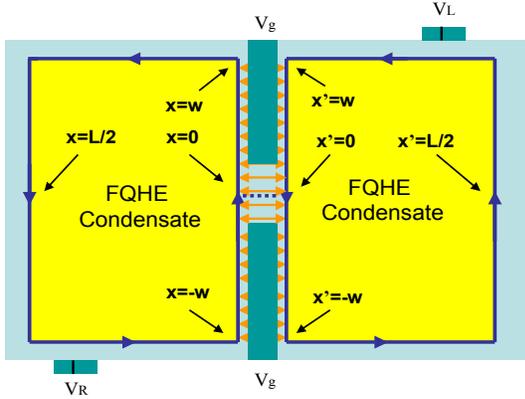}}
\end{picture}
\end{center}
\caption{Schematic illustration of a Hall bar with a closed gate defined constriction. 
The number of electrons in each quantum Hall system is a good quantum number in the absence of tunneling, which we treat perturbatively.
We assume that each Hall bar has perimeter $L$ and that the edges 
systems interact only along segment of length $2w$ along the gate.
}
\label{fig:3a}
\end{figure}

We start this section by comparing our model with more realistic models 
of split Hall bar systems with a line junction.  In the experimental 
geometry of the Pisa group \cite{pellegrini}, the barrier width is of order
of ten magnetic lengths and a barrier length at least forty times longer.  
At low gate voltages, tunneling across the line junction barrier is found experimentally
to be relevant even for $\nu<1$.  This property is surprising from the point of view of 
simple models of a Hall bar with constrictions.
A number of ideas have been advanced as possible explanations for this behavior\cite{vignale,eggert,halperin}
including the idea\cite{papa} that it is due to repulsive interactions across the junction.
Indeed with the Pisa geometry, interactions across the junction between the two
regions will always be important\cite{papa} unless screening by a gate is more effective for interactions
across the gate than for interactions on the same side of the gate.  As explained in the previous section,
the low-energy fixed point of our model is a split Hall bar if the top-bottom 
quasiparticle tunneling at the constriction has scaling dimensions
$d=\nu [(1+g_-)/(1-g_-)]^{1/2}<1$ and a single Hall bar otherwise. 
To illustrate the behavior of a more realistic model, we consider the case of a 
split Hall bar with two incompressible 
regions and allow arbitrary interactions between 
the charge densities at any two points along the edges.  
We limit our attention here to examining the relevance of tunneling of electrons across the line 
junction in this case. It is expected \cite{pryadko} that left-right electron tunneling in
the closed constriction case is dual to the top-bottom inter-edge
quasiparticle tunneling of the open constriction limit Fig.~\ref{fig:one} under fairly general circumstances; very similar realistic 
calculations could be conducted starting from the joined Hall bar fixed point and would be expected to lead to similar conclusions.  
Therefore we examine only the case of the closed constriction

In the closed constriction geometry tunneling is realized not by quasi-particles, but by electrons. 
Hence the electron tunneling operator has the form 
$\hat{T}_{\rm tunn}^{\rm el}(x=0,t)=R^\dagger(0,t) L(0,t) + {\rm H.c}$.
We have evaluated the tunneling correlation function using a numerical method that can be applied as 
easily for any edge-density edge-density interaction model.  
We report on results for the case $L=1200l_B$.
The edge lengths are chosen to be equal and we include additional interactions
only in a region of length $2w$ around the gates.  The model we study is therefore a more 
realistic version of the toy model.   
Our results do not depend on the actual edge perimeter length but on the ratio of the inter-edge interaction 
region length $w$ to length $L$ (it is assumed the edge length is long enough to overcome the 
discretized quantization of the conductance).

\begin{figure}
\unitlength=1mm
\begin{center}
\begin{picture}(80,55)
\put(2,-3){\includegraphics[height=50\unitlength,width=70\unitlength]{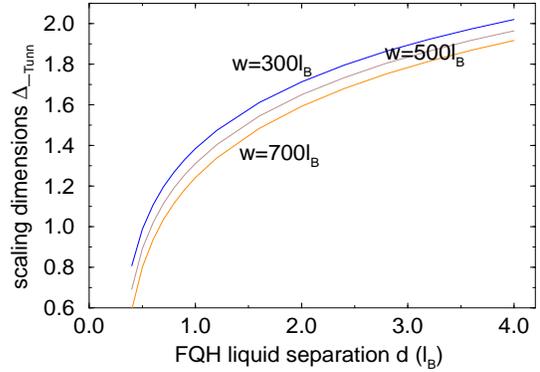}}
\end{picture}
\end{center}
\caption{Scaling dimensions of the tunneling operator between left and right FQH liquids as a function
of the distance between the two subsystems across the barrier.
These results are for edge perimeter $L=1200l_B$, $w=L/4=300l_B$ and $\nu_L=\nu_R=1/3$.
In the case when the separation of the layers is large $d\gg l_B$ the
scaling dimension approaches $1/\nu$ and tunneling is irrelevant. 
(w in the figure is used for $2w$ of text). }
\label{fig:three}
\end{figure}

The action for this system is
\bea
S &=& \sum_{\lambda} \int_0^{L}\rd x \rd \tau
      \frac{i}{2\pi \nu_\lambda}
      \p_x \phi^\lambda(x,\tau) \p_\tau \phi^\lambda(x,\tau)
\nonumber\\[1mm] \nonumber
&+& \hspace{0mm}
      \sum_{\lambda,\mu} \int_0^{L}  
                         \rd x \rd x' \rd \tau
       \p_x \phi^\lambda(x,\tau) H^{\lambda \mu}(x,x') \p_{x'}\phi^\mu(x',\tau)  \; ,
\\*
\eea
where $\lambda=R,L,$ and same for $\mu$. $H^{\lambda \mu}$ is the charge-charge density interacting kernel.
It is composed of intra- and inter-edge interaction.   
For the former we use the bare Coulomb interactions
\bea
V_{\rm RR}(x) = V_{\rm LL}(x) =  \frac{e^2}{\epsilon \sqrt{x^2 + \alpha l_B^2}} \quad.
\label{potential_same}
\eea
The interedge interaction $V_{\rm RL}$ is also modeled as Coulomb interaction, 
except that $l_B$ is substituted by the distance between edges $d$ as a short-distance cutoff.
Interactions outside of the interaction region of width $2w$, centered around $x=x'=0$, are neglected.
In Eq.~(\ref{potential_same}) $\alpha$  is dimensionless constant measure of short distance cut-off of order unity
whereas $\epsilon$ is the host semiconductor
dielectric constant which for the ${\rm GaAs/Al_{0.15}Ga_{0.85}As }$ heterojunction is $\epsilon=12.6$.

The electron tunneling-tunneling correlation function is given by Eq.~(\ref{tunn_correl2}) 
with constituent operators
\begin{equation}
G_{R^\dagger L^\dagger}(t,t') = \left<R^\dagger(t) L^\dagger(t')\right> = 
                                  \bigl<e^{i\frac{\sqrt{4\pi}}{\nu_R}\phi_R(t)}
                                  e^{-i\frac{\sqrt{4\pi}}{\nu_L}\phi_L(t')} \bigr>,
\label{el_CF}
\end{equation}
up to numerical factors.  We calculate the tunneling correlation function from the 
following combination of phase field correlation functions: $\left< \phi^\lambda(x=0;\tau) \phi^\mu(x'=0;\tau')\right>$,
\begin{equation}
\sum_{\lambda,\mu}   \frac{1}{\nu_\lambda\nu_\mu} 
        \left<2 \phi_\lambda(\tau) \phi_\mu(\tau')  \right> - \frac{1}{\nu^2_\lambda}\left<\phi^2_\lambda(\tau) 
        \right> - 
        \frac{1}{\nu^2_\mu}\left<\phi^2_\mu(\tau')  \right> \,,
\ee
which can be evaluated from the action by performing a numerical Bogoliubov transformation:  
\bea
&&\hspace{-2mm}\sum_{\lambda,\mu} \frac{1}{\nu_\lambda \nu_\mu }
                     \left< \phi^\lambda(0,\tau) \phi^\mu(0,\tau')\right>
 \hspace{-1mm}= \hspace{-2mm}\sum_{\lambda,\mu} \frac{1}{\nu_\lambda \nu_\mu }
\hspace{-1mm}\int \hspace{-1mm}  \frac{\rd^2 q}{(2\pi)^2} \frac{\rd \omega}{2\pi} 
\nonumber\\[1.5mm] \nonumber
&&
  \hspace{30mm}   \times  e^{i\omega(\tau-\tau')}      \left< \phi^{*\lambda}(q,\omega) \phi^\mu(q',\omega)\right>  ,
\\[3.5mm]\nonumber
&&= \sum_{\lambda,\mu} \frac{1}{\nu_\lambda \nu_\mu }
\int    \frac{\rd^2 q}{(2\pi)^2} \frac{\rd \omega}{2\pi} 
        e^{i\omega(\tau-\tau')}
\left(A^{-1}\right)^{\lambda \mu}(q,q';\omega) \; ,
\\[3.5mm]\nonumber
& =&  \hspace{-2mm} \sum_{\lambda,\mu} \frac{1}{\nu_\lambda \nu_\mu }
\int_{q,\omega}  \hspace{-2mm}
                e^{i\omega(\tau-\tau')}
\left[\frac{1}{2} M ( d^{-1} + d^{*-1})M^\dagger\right]^{\lambda \mu}\hspace{-4mm}(q,q';\omega) \, .
\nonumber\\*
\eea 
In these equations, the matrices $M$ and $M^\dagger$ that diagonalize the action
satisfy $MJM^\dagger =J$, and are found numerically, and
$\lambda_{q_l}$ is an eigenvalue of $JH$.

In Fig.~\ref{fig:three} we summarize our numerical results for the tunneling operator scaling dimension,
for the model described above and bulk filling factor $\nu=1/3$. 
Tunneling between subsystems becomes relevant even for $\nu=1/3$ when the  
two subsystems are separated by about one magnetic length. 
We expect that interactions across the junction will always be important in 
determining transport properties of cleaved-edge-overgrowth
line junction systems\cite{kang}, 
and that they can be important in split gate line junction systems, depending on the out-of-plane
distance to the nearby metallic gates.  
In addition to screening effects, these numerical calculations do not take into account
a number of other factors that could alter the ratio of the inter- to intra-edge
interactions. 
As can be seen in the Fig.~\ref{fig:three}, by increasing the length of the interaction region $w$, 
the scaling dimensions of left-right tunneling process decrease. 
If we would assume that the non-coplanar gates create smoother edges under the gate than in the rest of the 
system then presence of disorder along the edges in this part effectively makes
the inter-edge interacting regions much longer than the gate length\cite{pryadko2}.
These numerical results demonstrate that the consequences of across the line junction are 
qualitatively captured by our split Hall bar model.  

In closing we comment on our use of a model with a single chiral channel along 
the edge, which is certainly oversimplified.  In general the number of channels 
present at the edge depends on microscopic considerations.  In the case of 
the fractional quantum Hall effect, single channel edges are possible only\cite{macdonald90}
at the Laughlin filling factors $\nu=1/m$, but even at these bulk filling factors 
in general the number of edge channels and their character depends on the microscopic details.  
The number of channels at an edge increases as the two-dimensional electron system
confining potential gets smoother, through a process known as edge reconstruction\cite{chklovskii,zulicke}.
In the limit of a smooth edge, a Tomas-Fermi approximation for the edge density profile becomes
accurate as the number of discrete microscopic channels becomes very large.  In this limit, 
the edge may be described as consisting of incompressible strips at a series of 
integer and fractional filling factors\cite{chklovskii} separated by compressible strips 
that correspond to a large number of one-dimensional channels.  
Gate defined edges of two-dimensional electron systems in particular, in a Tomas-Fermi approximation, 
have a characteristic \cite{larkin} square root density-profiles with the distance to the gate $a$ as 
a characteristic length scale.  In ignoring this complex behavior we are implicitly appealing to 
the overriding strength of Coulomb interactions which will tend to create one high-velocity 
which corresponds to a rigid shift of the edge and which interacts relatively weakly with the 
(possibly many) lower velocity charge-neutral internal edge excitations.  There is considerable 
experimental evidence from tunneling experiments\cite{changreview} that this pragmatic expedient 
has substantial validity in many circumstances.  
In general these extra {\em phonon} edge modes
can\cite{eggert,halperin} change the scaling dimensions of the correlation function of the edge fields,
tending to produce orthogonality catastrophe's that tend to make tunneling processes less relevant. 
The quantitative importance of this effect is, however, difficult to estimate.  

In summary, this paper discusses the properties of a simple analytically solvable model
that captures the principle geometric effects responsible for the lack of universality of 
quantum Hall edge transport properties.  In particular, translational invariance along the edge
does not hold in typical experimental geometries, with strong interactions possible either across the 
Hall bar or across a junction in the Hall bar that enhances back scattering, even between points 
that are remote when distance is measured along the edge. 
The properties of this model help to explain why simple Hall bar backscattering models\cite{kane,fendley}
are not able to account for many aspects of the experimental data.
We believe that our model is especially relevant for systems with long-thin geometries
where interactions across a barrier ($g_1$) 
and across a Hall bar ($g_2$) can both be important
and tuned experimentally\cite{kang,grayson,kane2}.

\section*{ACKNOWLEDGMENTS}

Some of these results were described briefly in a previous publication.\cite{papa}
This work was supported by the Welch Foundation and by the National Science Foundation under grant 
DMR-0115947. The authors are grateful to Dave Allen, Joseph Betouras, Vittorio Pellegrini, Leonid Pryadko,
Stefano Roddaro, Yun-Pil Shim, Tilo Stroh, Giovani Vignale and most of all to 
Alexei Tsvelik for valuable discussions and comments.

\appendix*{}

\section{}
\label{append}

\subsection{Tunneling-tunneling correlation function}
 We make use in the text of the identity 
\be
\label{campbell}
\left<\prod_{i} e^{\hat{O}_i} \right> = 
\exp\left(\frac{1}{2}\left<\left[2\sum_{i< j} \hat{O}_i \hat{O}_j + \sum_i \hat{O}^2_i   \right]\right>\right) .
\ee
Applying this formula and using the representation of the electron creation 
operator as $R \sim e^{i \phi(x)}$ in the tunneling operators 
$T_{\rm tunn}^{\rm el}(x,x';t)=R^\dagger(x,t) R(x',t)$ the required correlation functions take the form
\bea
\label{tunn_correl3}
\bigl<\hat{T}_{\rm tunn}^{\rm el}(x,x';t) \hat{T}_{\rm tunn}^{\rm el}(x'',x''';t')\bigr> \hspace{-2mm}&=& \hspace{-2mm}
\frac{G(x,x';t,t) G(x,x''';t,t')}{G(x,x'';t,t')}
\nonumber \\[3mm]
&& \hspace{-20mm}\times \frac{G(x',x'';t,t') G(x'',x''';t',t')}{G(x',x''';t,t')} ,
\eea
where $G(x,x';t,t')=\left<R^\dagger(x,t) R(x',t') \right>$.

\subsection{Calculation of sums $S_{\rm e/o}(q_1,\Delta z)$}
We give here some details of the calculation of the sums $S_{\rm e/o}(q_1,\Delta z)$ that appear in the text
\begin{equation}
\label{phase_sum2}
S_{\rm e/o}(q_1,\Delta z)  \hspace{-1.mm} = \hspace{-2mm}\sum_{n>0}^{\rm e/o} \frac{1}{4\pi n}
\left[\left(e^{iq_{n} \Delta z} -1\right) \bar{N}_{n} +
\left(e^{-iq_{n}\Delta z} -1\right)N_{n} \right]
\end{equation}
where $q_n=nq_1$ with $q_1=2\pi/L$. 
We write the phase fluctuating terms $(e^{\pm i n q_1 \Delta z}-1)$ as 
$\cos(nq_1\Delta z)-1 \pm i \sin(nq_1\Delta z)$ and notice that the sum of their imaginary parts is multiplied 
by $(\bar{N}_n - N_n)=1$ whereas the sum of their real parts is multiplied by $(\bar{N}_n + N_n) = 1 + 2 N_n$,
where
\bea
N_n = \frac{1}{e^{q_1{v}n \beta}-1} \quad, \quad 
\bar{N}_n = \frac{1}{1-e^{-q_1{v}n \beta}} \quad. 
\eea
Therefore the sums $S_{e/o}(q_1,\Delta z)$ take the form
\bea
S_{{\rm \rm e/o}}(q_1,\Delta z) & = &
- \frac{1}{\pi}\sum_{n>0}^{\rm e/o}\frac{1}{n}\frac{e^{-nq_1v\beta}}{1-e^{-nq_1v\beta}}
\sin^2(nq_1\frac{\Delta z}{2})
\nonumber\\[2mm]
&& + \frac{i}{4\pi}\sum_{n>0}^{\rm e/o}\frac{1}{n}\sin(nq_1\Delta z)
\nonumber\\[2mm]
&& -
\frac{1}{2\pi}\sum_{n>0}^{\rm e/o}\frac{1}{n}\sin^2(nq_1\frac{\Delta z}{2}) \quad .
\eea
The second term is divergent and needs to be regularized.  We do this by multiplying with an exponent
$e^{-\epsilon n}$ and take again the limit $\epsilon \to 0$, using Eq.~(1.462) in Ref.~\onlinecite{gradshteyn}.
For the third term
we can combine the Eqs.~(16.30.1) and (16.30.2) of Ref.~\onlinecite{stegun} to get
\bea
S_{\rm e}(q_1,\Delta z) & = &
 \frac{1}{2}\frac{i}{8\pi}\Bigl(\pi-(2 q_1 \Delta z){\rm mod} (2\pi)\Bigr)
\nonumber \\[3mm]\nonumber
&& - \frac{1}{8\pi}\ln\left[
\frac{ \vartheta_1(q_1\Delta z/2, e^{-q_1 v\beta/2})}
{\vartheta_1'(0,e^{-q_1 v\beta/2})\sin(q_1\Delta z)}
\right.
\\[3mm]\nonumber
&&
\left.
\times \sqrt{1+\frac{\sin^2(q_1\Delta z)}{\sinh^2(q_1\alpha/2)}} \right]
\\[3mm]
&& - \frac{1}{8\pi}\ln\left[
\frac{\vartheta_2(q_1\Delta z/2, e^{-q_1 v\beta/2})}
{\vartheta_2(0,e^{-q_1 v\beta/2})} \right] ,
\eea
where $\vartheta_i$ is an elliptic theta function.
We substitute below
$\vartheta'_1(0,e^{-q_1 v\beta/2})=
\vartheta_1(\tilde{\alpha}-\tilde{\beta},e^{-q_1 v\beta/2})/ \sin(\tilde{\alpha}- \tilde{\beta})$,
for $\tilde{\alpha}  \rightarrow \tilde{\beta}$.
We assume here that $\alpha \to 0$ for $q_1\Delta z\to 0$ is done such
that $\alpha/(q_1 \Delta z) \to 0$ therefore $S(q_1,\Delta z)$ can be written as in the text.

For the odd components we have to subtract Eqs.~(16.30.1) and (16.30.2) of Ref.~\onlinecite{stegun}
to obtain Eq.(~\ref{odd_sum}).
In the text we use also the abbreviations $F(\Delta z)$ and $\tilde{F}(\Delta z)$ for the following ratios of 
$\vartheta_1$ and $\vartheta_2$ functions
\bea
  F(\Delta z) & = &
     {\rm e}^{{\rm i}(\frac{\pi\Delta z}{L}-\frac{\pi}{2})}
\frac{ \vartheta_1\left(\frac{\pi \Delta z}{L}|{\rm i}\frac{v\beta}{L}\right)  }
     { \vartheta_1\left(\frac{\pi \alpha}{L}|{\rm i}\frac{v\beta}{L} \right)   }
   \; ,\;
\eea
\bea
\tilde{F}(\Delta z) & = &
     {\rm e}^{{\rm i}(\frac{\pi\Delta z}{L})}
\frac{ \vartheta_2\left(\frac{\pi \Delta z}{L}|{\rm i}\frac{v\beta}{L}\right)  }
     { \vartheta_2\left(\frac{\pi \alpha}{L}|{\rm i}\frac{v\beta}{L} \right)   }
  \,.
\eea

\subsection{Conductance of finite size systems}

Here we show some details of the calculation of the conductance of systems of finite size.
To this end we make use of the binomial expansion:
\be
  (1-a)^{-2d}=\sum_{n=0}^\infty \textbinm{-2d}{n}(-a)^n=
  \sum_{n=0}^\infty \textbinm{2d +n-1}{n} a^n \;.
  \label{eq:bin-exp}
\ee
The tunneling current is
\begin{equation}
I^{\rm QP}_{\rm tunn} = \left[1-\exp\left(-\frac{eV}{k_\abbr{B}T}\right)\right]
 \frac{e^*}{\hbar} \int_{-\infty}^{+\infty} \rd t \, \eun ^{\imu eVt} \;
 G^{\rm QP}_{\rm tunn}(t)\; ,
\label{I_tunn_2}
\end{equation}
with the Fourier transform of correlation function being
\bea
&&\hspace{-3mm}\int_{-\infty}^{+\infty} 
\rd t \, \eun ^{\imu eVt}
\left[
\imu\frac{L}{2\pi}\sin\left(\frac{2\pi v_-t}{L}\right)
\right]^{-2d}
\nonumber \\[3mm]\nonumber
&& = \left(\frac{L}{4\pi}\right)^{-2d} 
\hspace{-2.5mm}  \int
\hspace{-1mm}  \rd t \,\eun^{\imu eVt} \eun^{-\imu\frac{4\pi d v_- t}{L}}
 \sum_{n=0}^\infty \textbinm{{2d +n-1}{n}} \eun^{-\imu n \frac{4\pi v_- t}{L}} ,
\\[3mm]\nonumber
&& =
\left(\frac{L}{2\pi}\right)^{-2d} \sum_{n=0}^\infty \textbinm{{2d +n-1}{n}}
2\pi \delta\left[eV-\frac{4\pi v_-}{L}(n+d)\right] .
\nonumber\\*
\eea
The sum in the second line needs to be regularized (e.g., by
$\eun^{-\epsilon n}$) because $a=e^{-i\pi z/L}$, $|a|=1$ in the
expansion (\ref{eq:bin-exp}). Writing the binomial coefficient in terms
of Gamma functions, $[\Gamma(2d +n)/\Gamma(n+1)\Gamma(2d)]$,
and using Sterling's formula 
for large system sizes,
\bea
\Gamma(z) \sim \eun^{-z} z^{z-\frac{1}{2} } (2\pi)^{\frac{1}{2}} + \ldots \quad ,
\eea
we find
\bea
{{2d +n-1}\choose {n}} \sim n^{2d - 1}  \quad ,
\eea
which is  in agreement with the dependence of the tunneling current on bias voltage for
infinite system sizes.


\begin{thebibliography}{50}
\bibitem{low_V}
We make use here of these identities
$B(x,y) =\Gamma(x)\Gamma(y)/\Gamma(x+y)$ and $\Gamma(z) \Gamma(1-z) = \pi/\sin(\pi z)$,
and\cite{gradshteyn}
\bea
\left| \frac{\Gamma(x+iy)}{\Gamma(x)} \right| = \prod_{n=0}^{\infty}\left[1+\frac{y^2}{(x+n)^2}
\right]^{-1} \quad.
\eea
\end{thebibliography}
\end{document}